%% file: ms.tex
\documentclass[prd,twocolumn,nofootinbib,superscriptaddress]{revtex4-1}

\usepackage{graphicx}
\usepackage{amssymb}
\usepackage{amsmath}
\usepackage{epstopdf}
\usepackage{aas_macros}
\usepackage{color}

\newcommand{\cpm}{{\small CUBEP$^3$M}}
\newcommand{\hfit}{{\small HALOFIT}}

\newcommand{\class}{{\small CLASS}}

\newcommand{\mpch}{\mbox{Mpc}/\mbox{h}}
\newcommand{\hmpc}{\mbox{h}/\mbox{Mpc}}

\newcommand{\mev}{{\rm meV}}

\begin{document}
\title{Cosmic neutrinos: dispersive and non-linear}
\author{Derek Inman}
\email{inmand@cita.utoronto.ca}
\affiliation{Canadian Institute for Theoretical Astrophysics,
University of Toronto, 60 St. George St., Toronto, ON M5S 3H8, Canada}
\affiliation{Department of Physics, University
of Toronto, 60 St. George, Toronto, ON M5S 1A7, Canada}
\author{Ue-Li Pen}
\email{pen@cita.utoronto.ca}
\affiliation{Canadian Institute for Theoretical
  Astrophysics, University of Toronto, M5S 3H8, Ontario, Canada}
\affiliation{Dunlap Institute for Astronomy and Astrophysics,
  University of Toronto, Toronto, ON M5S 3H4, Canada}
\affiliation{Canadian Institute for Advanced Research, Program in
  Cosmology and Gravitation} \affiliation{Perimeter Institute for
  Theoretical Physics, Waterloo, ON, N2L 2Y5, Canada}

\input{./abstract.tex}

\maketitle

\input{./introduction.tex}

\input{./theory.tex}

\input{./results.tex}

\input{./discussion.tex}

\input{./conclusion.tex}

\input{./acknowledgements.tex}

\bibliographystyle{apsrev}
\bibliography{thebib}
\widetext
\clearpage
\input{./supplement.tex}

\end{document}

%% file: abstract.tex
\begin{abstract}
  We present a description of cosmic neutrinos as a dispersive fluid.
  In this approach, the neutrino phase space is reduced to density and
  velocity fields alongside a scale-dependent sound speed.  This sound
  speed depends on redshift, the initial neutrino phase space density
  and the cold dark matter gravitational potential.  The latter is a
  new coupling between neutrinos and large scale structure not
  described by previous fluid approaches.  We compute the sound speed
  in linear theory and find that it asymptotes to constants at small
  and large scales regardless of the gravitational potential.  By
  comparing with neutrino N-body simulations, we measure the small
  scale sound speed and find it to be lower than linear theory
  predictions.  This allows for an explanation of the discrepency between
  N-body and linear response predictions for the neutrino power
  spectrum: neutrinos are still driven predominantly by the cold dark
  matter, but the sound speed on small scales is not stable to
  perturbations and decreases.  Finally, we present a calibrated model
  for the neutrino power spectrum that requires no additional
  integrations outside of standard Boltzmann codes.
\end{abstract}

%% file: introduction.tex
\begin{section}{Introduction}
  \label{sec:introduction}
  Neutrinos are an important part of both the Standard Model of
  Particle Physics and the Standard Model of Cosmology.  However, many
  of their properties, such as mass and chirality, have yet to be
  determined. One way to probe neutrinos is through large scale
  structure surveys which measure tracers of the density field.  The
  principal effect of neutrinos on the density field is a suppression
  of the total matter power (including CDM, baryons and neutrinos) on
  small scales caused by the fast thermal motions of neutrinos.

  Since CDM gravitational dynamics are very non-linear, simulations
  including neutrinos must be performed.  A variety of strategies have
  been used to include neutrinos.  The most accurate is to include
  them as a separate N-body particle \cite{bib:Viel2010}.  However,
  due to the Poisson noise from their thermal motions, many neutrino
  particles are needed and most of the simulation memory is used in
  storing neutrinos, despite their small effects.  A vastly more
  efficient way is to treat neutrinos as a linear response to the cold
  dark matter (CDM) and only compute their transfer function at each
  timestep \cite{bib:AliHaimoud2012}.  While this approach correctly
  obtains the suppression in power, there is a consistent
  deficit seen in the neutrino power spectrum compared to the N-body
  simulations.  This deficit occurs even though the CDM is computed
  fully non-linearly indicating that neutrinos are not accurately
  described by first order perturbation theory.

  A variety of approaches have been developed to treat neutrinos
  beyond linear theory.  In \cite{bib:Fuhrer2014}, the Vlasov equation
  is perturbatively expanded to include higher order contributions.  A
  novel approach is described by \cite{bib:Dupuy2013} who utilize the
  fact that neutrinos are collisionless to describe them as a set of
  many non-interacting flows.  Finally, in \cite{bib:Massara2014}, the
  authors use numerically determined neutrino halo profiles to compute
  the one-halo contribution to the neutrino power spectrum.

  In this work we consider to what degree neutrinos can be described
  as a {\it dispersive} fluid, i.e. one where the sound speed varies
  with wavenumber.  Fluid approaches are advantageous as they reduce
  the high dimensionality of neutrino phase space to a smaller and
  more managable set of hierarchy equations.  In other words, the
  velocity distribution of particles need not be evolved.  Due to
  their simplicity, there are many studies of non-dispersive fluids,
  e.g. \cite{bib:Shoji2010} (as well as many others).  In a
  non-dispersive fluid, the sound speed is independent of the CDM
  perturbations which drive, but are not directly coupled to, the
  neutrino perturbations. Here we demonstrate that a dispersive
  approach is required even for the linearized Vlasov solution making
  it useful to study.  Furthermore, we demonstrate that the N-body
  neutrino power spectrum can be accurately reproduced through
  straightforward modifications of the small scale sound speed.

  The simulations used in this paper are the same as in
  \cite{bib:Inman2015} where neutrinos are implemented as a distinct
  N-body particle into the \cpm{} code \cite{bib:HarnoisDeraps2012}.
  In order to discuss time-dependence, we have run simulations with
  the same code but half the size (per dimension): $64$ nodes instead
  of $512$ and a cubic volume with lengths of $250$ instead of $500$
  \mpch{}.  For $m_\nu=50$ \mev{}, we use power spectra from the
  significantly larger TianNu simulation \cite{bib:Yu2016} in order to
  better deal with neutrino Poisson noise.  We note that the power
  spectrum was computed slightly differently in this work as it used
  NGP particle interpolation and did not use the groups method
  described in \cite{bib:Inman2015}.  Instead, we simply subtract the
  predicted shot noise power spectrum from the neutrino power.

  We often need to integrate against a source potential, $\phi$, for
  which we use the \class{} code \cite{bib:Blas2011}.  We always use
  Poisson's equation to change $\phi\rightarrow\delta_m$ and then
  replace $\delta_m$ by the matter transfer function, $T_m(k)$,
  outputted by \class{}, including neutrinos, and with the non-linear
  correction $T_m\rightarrow T_m\sqrt{P_{NL}/P_{L}}$ where $P_L$ is
  the linear power spectrum and $P_{NL}$ is the non-linear \hfit{}
  also outputted by \class{}.  We note that the simulations were
  normalized to have the same $\sigma_8$ which could yield small
  discrepencies between \class{} and N-body results.

\end{section}

%% file: theory.tex
\begin{section}{Theory}
  \label{sec:theory}
  \begin{subsection}{Vlasov Equation}
    \label{ssec:vlasov_eqn}
    The Vlasov equation in an expanding Universe for non-relativistic
    particles well inside the Hubble scale is given by
    \begin{align}
      \label{eq:vlasov_eqn}
      f_s + v^i f_{x^i} - a^2 \phi_{x^i} f_{v^i} = 0 
    \end{align}
    where subscripts denote partial differentiation, $a$ is the
    scalefactor, $s$ is the Newtonian (``Superconformal'') time
    defined by $dt = a^2 ds$, $v^i= a \frac{dx^i}{d\tau}$ is the
    conjugate velocity with $d\tau = a dt$ being the conformal time
    and $x^i$ being the comoving position, $f$ is the one particle
    distribution function and $\phi$ is the gravitational potential.
    For a pedagogical discussion of this equation we refer the reader
    to \cite{bib:Bertschinger1995}.  $\phi$ can be computed from the
    matter field via Poisson's equation:
    \begin{align}
      \phi_{x^ix^i} = 4 \pi G \rho_{cr} \delta_m a^2 = \frac{3}{2} H_0^2
      \Omega_m \frac{\delta_m}{a} \nonumber
    \end{align}
    where $\rho_{cr}$ is the critical density of the universe,
    $\delta_m$ is the matter density contrast defined via
    $\rho_m = \rho_{cr}(1+\delta_m)$, $H_0$ is the present day Hubble
    parameter and $\Omega_m = \Omega_c + \Omega_b + \Omega_\nu$ is the
    present day matter fraction of the Universe.  Since
    $\Omega_\nu \ll 1$, $\phi$ is approximately independent of
    neutrinos and Eq. \ref{eq:vlasov_eqn} is linear in neutrino
    perturbations.  Nonetheless, it is not first order in cosmological
    perturbations until it is ``linearized'' by taking
    $f_{v^i} \rightarrow f^0_{v^i}$ with $f^0(v; \beta)$ being the
    relativistic Fermi-Dirac distribution:
    \begin{align}
      \label{eq:fermi_dirac}
      f^0(v ; \beta) = \frac{1}{e^{\beta v}+1} \nonumber\\
      \bar{f}^0(v) = f^0(v; 1)
    \end{align}
    with $\beta = \frac{m}{k_B T_\nu c}$ and $\bar{f}^0$ will be used
    in subsequent calculations.  This is equivalent to neglecting the
    acceleration,
    $\frac{\partial v}{\partial s} = -a^2 \phi_{x^i} \simeq 0$,
    leading to the term ``free streaming''.  Furthermore, it adds a
    source term given by a homogeneous background of neutrino
    particles.  The integral solution to this equation is easy to
    obtain in Fourier space as:
    \begin{align}
      \label{eq:vlasov_sln}
      f(s,\vec{k},\vec{v}) = &f(s_i,\vec{k},\vec{v}) e^{-ik^iv^i(s-s^i)}
                               + \nonumber \\ &\int_{-\infty}^s ds' a(s')^2 i k^i \phi(s',k) f^0_{v^i}(v) e^{-i
                                                k^iv^i(s-s')}.
    \end{align}
    This solution has been used in many works of which we reference a
    few more modern ones
    \cite{bib:Ringwald2004,bib:Shoji2010,bib:AliHaimoud2012}. One can
    then compute expectation values of the distribution,
    $\langle A \rangle = \int d^3v A f / \int d^3v f^0$ which give
    quantities like the density contrast,
    $\delta = \langle f \rangle $, and the divergence of the stress
    tensor, $\Pi = i^2 k^ik^j \langle v^i v^j \rangle$.  For
    Eq. \ref{eq:vlasov_sln}, we derive in the Supplement the
    following:
    \begin{align}
      \delta &= \int ds' a^2(-k^2\phi) (s-s')
               \langle j_0(k u (s-s')/\beta) \rangle_0 \label{eq:vlasov_den} \\
      -\frac{\beta^2}{k^2}\Pi &= \int ds' a^2(-k^2\phi)(s-s')
                                \langle u^2 j_0(k u (s-s')/\beta) \rangle_0 \label{eq:vlasov_str} 
    \end{align}
    where
    $\langle F(x,u) \rangle_0 = \int u^2 \bar{f}^0(u) F(x,u) du / \int
    u^2 \bar{f}^0(u) du$,
    $j_0(x) = \sin(x)/x$ is the first order spherical bessel function
    and we have changed velocity variables to the dimensionless
    $u=\beta v$.
  \end{subsection}

  \begin{subsection}{Moment Equations}
    \label{ssec:moment_eqn}
    An alternative approach to solving Eq. \ref{eq:vlasov_eqn} is to
    derive differential equations for the moments themselves.  In the
    context of neutrinos, the fluid approximation is studied in detail
    by \cite{bib:Shoji2010} and also in the \class{} paper
    \cite{bib:Lesgourgues2011}.  The first two moments yield the
    continuity and Euler equations:
    \begin{align}
      \delta_s + \theta &= 0 \nonumber \\
      \theta_s + \Pi &= a^2 ( (1+\delta)\phi_{x^i})_{x^i} \nonumber 
    \end{align}
    which can be combined by eliminating $\theta$ and introducing the
    sound speed $\Pi = -c_s^2 k^2 \delta$:
    \begin{align}
      \label{eq:fluid_eqn}
      \delta_{ss} + c_s^2 k^2 \delta = a^2 (-k^2 \phi)
    \end{align}
    where we have linearized the right hand side.  In the Supplement
    we compute the Green's function solution, assuming $c_s$ is
    constant, to this equation and find:
    \begin{align}
      \label{eq:fluid_sln}
      \delta_{c_s} &= \int ds' a^2(-k^2\phi) (s-s') j_0(k c_s (s-s')) \\
      \Pi &= - c_s^2 k^2 \delta_{c_s} \nonumber
    \end{align}
    where we have added the subscript $_{c_s}$ to differentiate from
    the Vlasov density contrast.  Comparing Eq. \ref{eq:fluid_sln} to
    Eq. \ref{eq:vlasov_den}, we see that by exchanging the order of
    integration we can re-write Eq. \ref{eq:vlasov_den} as
    \begin{align}
      \label{eq:fluid_sum}
      \delta = \frac{ \int u^2 \bar{f}^0(u) \delta_{ (u/\beta) } du}{\int
      u^2 \bar{f}^0(u) du},
    \end{align}
    that is, as a weighted sum of fluid solutions.  In principle this
    allows a measurement of the linear neutrino power to be decomposed
    into a sum of fluid solutions, the distribution of which yielding
    information on the neutrino velocity distribution.

  \end{subsection}

\end{section}

%% file: results.tex
\begin{section}{Results}
  \label{sec:results}

  \begin{subsection}{Sound speeds}
    In solving Eq. \ref{eq:fluid_eqn}, we assumed a constant sound
    speed despite the fact that $\Pi = -c_s^2 k^2 \delta$ does not
    enforce $c_s$ to be constant.  Since we have solutions for $\Pi$
    and $\delta$ in Eq. \ref{eq:vlasov_str} and \ref{eq:vlasov_den},
    we can compute the exact sound speed as a function of wavenumber:
    \begin{align}
      c_s^2  = \frac{1}{\beta^2}\frac{ \int_{-\infty}^{s} ds'
      a^2(-k^2\phi)(s-s') \langle u^2 j_0(ku(s-s')/\beta) \rangle_0 }{ \int_{-\infty}^{s} ds'
      a^2(-k^2\phi)(s-s') \langle j_0(ku(s-s')/\beta) \rangle_0 }.
    \end{align}
    We show this as solid lines in Fig. \ref{fig:vlasov_cs}.
    \begin{figure}[htbp]
      \begin{center}
        \includegraphics[width=0.5\textwidth]{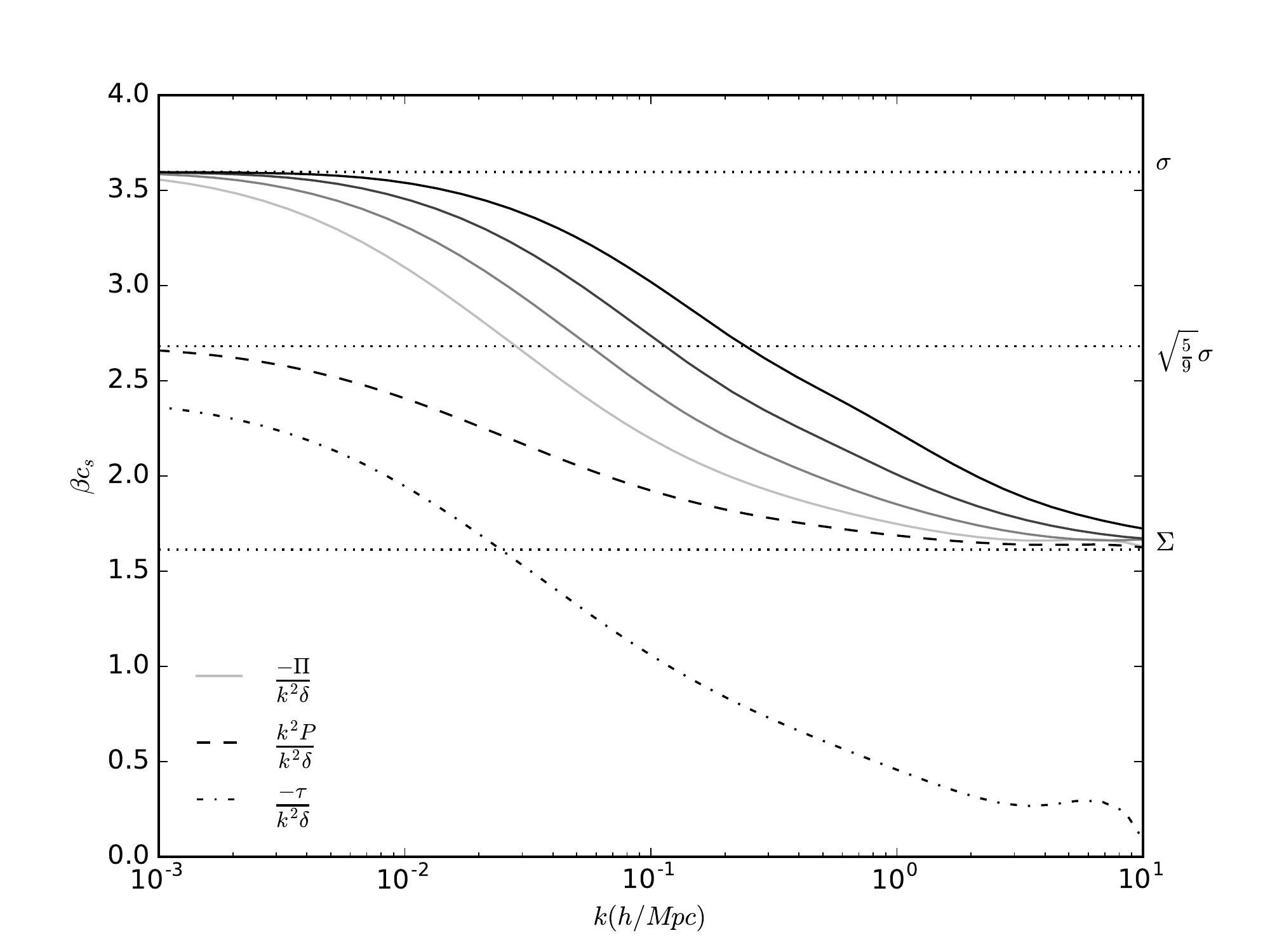}
        \caption{Sound speeds computed from the linearized Vlasov
          equation.  Solid lines are computed with respect to the
          stress , $\Pi$, with darker lines indicating heavier
          neutrino mass.  The dashed and dash-dotted lines are
          computed with respect to the pressure $k^2P$ and the
          anisotropic stress $\tau$ for $m_\nu = 50$ \mev.  Horizontal
          dotted lines are pre-computed asymptotic behaviours which
          are independent of both neutrino mass and time.
          $\beta=m/(k_BT_\nu c)$.}
        \label{fig:vlasov_cs}
      \end{center}
    \end{figure}
    We see that neutrinos are approximately bimodal with constant
    sound speed at large and small $k$.  Both these values are
    computable.  For $k\rightarrow 0$, $j_0(ku(s-s')/\beta) \simeq 1$.
    This means the velocity integral becomes separable from the $s$
    integral and we find
    \begin{align}
      \label{eq:cs_lowk}
      -\frac{\beta^2}{k^2}\Pi &\simeq 
                                \int ds' a^2(-k^2\phi)(s-s')
                                \sigma^2 \nonumber \\
      \delta &\simeq \int ds' a^2(-k^2\phi)(s-s')  \\
      \therefore  c_s^2 &=\left( \frac{\sigma}{\beta} \right)^2 \nonumber
    \end{align}
    where $\sigma$ is the velocity dispersion
    $\sigma^2 = \langle u^2 \rangle_0 \simeq 12.94$.  For large $k$,
    the sinusoids in the velocity integral oscillate rapidly and add
    to zero unless $s'\simeq s$
    \cite{bib:Ringwald2004,bib:AliHaimoud2012}.  Under this
    assumption, $(a\delta_m)(s') \simeq (a\delta_m)(s)$ and can be
    factored out of the integral yielding:
    \begin{align}
      \label{eq:cs_highk}
      -\frac{\beta^2}{k^2}\Pi &\simeq a^2(-k^2\phi) \left(\frac{\beta}{k}\right)^2 \nonumber \\
      \delta &\simeq a^2(-k^2\phi) \left(\frac{\beta}{k}\right)^2
               \Sigma^{-2}  \\
      \therefore c_s^2& =  \left(\frac{\Sigma}{\beta}\right)^2 \nonumber
    \end{align}
    where one must be particularly careful in changing the order of
    integration and $\Sigma$ is an ``inverse dispersion''
    $\Sigma^{-2} = \langle u^{-2} \rangle_0 \simeq 0.38$.  This second
    sound speed is also the one that goes into defining the free
    streaming wavenumber
    $k_{fs} = \sqrt{ \frac{3}{2} \Omega_m a }
    H_0\frac{\beta}{\Sigma}$.
    A simple ``instantaneous'' approximation on small scales is easily
    found by considering Eq. \ref{eq:fluid_eqn}: for large $k$,
    $\delta_{ss} \ll c_s^2 k^2 \delta$ and so
    $\delta \simeq \frac{-a^2k^2\phi}{c_s^2 k^2}$ or:
    \begin{align}
      \label{eq:delta_highk}
      \delta \simeq \left(\frac{k_{fs}}{k} \right)^2 \delta_m
    \end{align}
    on small scales.  Equivalently, we can treat this as an equation
    for the sound speed:
    \begin{align}
      \label{eq:cs_highk_den}
      c_s^2 = \frac{ \frac{3}{2}H_0^2 \Omega_m a}{k^2} \frac{\delta_m}{\delta}.
    \end{align}
    Finally, we can divide the stress tensor into pressure, $P$, and
    anisotropic stress,$\tau$, $\Pi = -k^2P + \tau$.  Equations for
    these components are again derived in the Supplement:
    \begin{align}
      \beta^2 P &= \int_{-\infty}^s ds' a^2 (-k^2\phi) (s-s') \frac{1}{3} \langle u^2
                  (j_0(ku(s-s')/\beta)+\nonumber \\ &\hspace{1cm}+2 j_1(ku(s-s')/\beta)/(ku(s-s')/\beta) \rangle_0 \label{eq:vlasov_pre} \\
      -\frac{\beta^2}{k^2} \tau &= \int_{-\infty}^s ds' a^2 (-k^2\phi) (s-s') \frac{2}{3} \langle u^2
                                  (j_0(ku(s-s')/\beta) \nonumber \\
                &\hspace{1cm}-j_1(ku(s-s')/\beta)/(ku(s-s')/\beta) \rangle_0 \label{eq:vlasov_ani}
    \end{align}
    We can now repeat our small scale approximations for these two
    components, e.g. $P/(\delta)$ and $\tau/(-k^2\delta)$.  For the
    pressure on small scales, we expand the sinusoides to find
    $\frac{1}{3} (j_0(x)+2j_1(x)/x) \simeq \frac{5}{9} x$ and so the
    sound speed is $\sqrt{5/9}\sigma$.  This result was derived by
    \cite{bib:Shoji2010}.  On small scales, repeating the above
    derivation shows that $k^2 P\rightarrow \Pi$ and
    $\tau\rightarrow 0$.  These approximations are shown as horizontal
    lines in Fig. \ref{fig:vlasov_cs} and closely match the integrated
    values (dashed and dash-dotted lines).
  \end{subsection}

  \begin{subsection}{Perturbed Distribution Function}
    We now repeat the arguments used in computing the large and small
    scale sound speed limits but for the distribution function
    instead.  Assuming negligible initial conditions,
    Eq. \ref{eq:vlasov_sln} can be written as:
    \begin{align}
      f(s,k,v,\mu) = \int_{-\infty}^s ds' a^2 i k \phi \mu
      \frac{df^0}{dv} e^{-ikv\mu(s-s')} \nonumber
    \end{align}
    where $\mu=\vec{k}\cdot\vec{v}/(kv)$.  We can now integrate over
    angles to find:
    \begin{align}
      \langle f(s,k,v) \rangle &= \frac{1}{2}\int_{-1}^1d\mu 
                                 f(s,k,v,\mu) \nonumber \\
                               &=\frac{df^0}{dv}\int_{-\infty}^{s} ds'
                                 a^2 k \phi j_1(kv(s-s')). \nonumber
    \end{align}
    In the large scale limit, $kv(s-s')\ll1$ and
    $j_1(kv(s-s')) \simeq kv(s-s')/3$.  Using this approximation and
    substituting the density obtained in Eq. \ref{eq:cs_lowk} yields:
    \begin{align}
      \langle f \rangle = -\frac{1}{3} v \frac{df^0}{dv} \delta. \nonumber
    \end{align}
    In the small scale limit limit limit, we again use $s\simeq s'$
    and find
    \begin{align}
      \langle f \rangle &= \frac{df^0}{dv} a^2 k \phi \int_{-\infty}^s
                          ds' j_1(kv(s-s')) \nonumber \\
                        &=\frac{1}{v}\frac{df^0}{dv} a^2 \phi
                          \int_0^\infty j_1(x) dx \nonumber \\
                        &=-\frac{1}{v}\frac{df^0}{dv}\frac{\Sigma^2}{\beta^2}
                          \delta \nonumber
    \end{align}
    where we use the density in Eq. \ref{eq:cs_highk} instead.  This
    result was also computed in \cite{bib:AliHaimoud2012} using a
    different technique.  We note now that strictly speaking these are
    not ``low-k'' and ``high-k'' limits, rather, they refer to limits
    where $kv\ll $ or $ \gg (\Delta s)^{-1}$ for some timescale
    $\Delta s$.  Nonetheless we will refer to the limits as such
    throughout the paper.  Both the low- and high-k perturbations are
    separable in position and velocity and so we can define the
    velocity space perturbation as
    $f^1(v) = \langle f \rangle (v,k,s)/\delta(k,s)$.  In terms of the
    dimensionless velocity $u=\beta v$ we have:
    \begin{align}
      \label{eq:perturbed_f}
      \bar{f}^1(u) &= \frac{\langle f \rangle (u,k,s)}{\delta(k,s)} \nonumber \\
                  &=\frac{1}{u}\frac{e^u}{\left ( e^u +1 \right )} \bar{f}^0(u) \begin{cases} \frac{1}{3}u^2
                    & ku(\Delta s)/\beta \ll 1 \\ \Sigma^2 & ku(\Delta s)/\beta \gg
                    1. \end{cases}
    \end{align}

    \begin{figure}
      \begin{center}
        \includegraphics[width=0.5\textwidth]{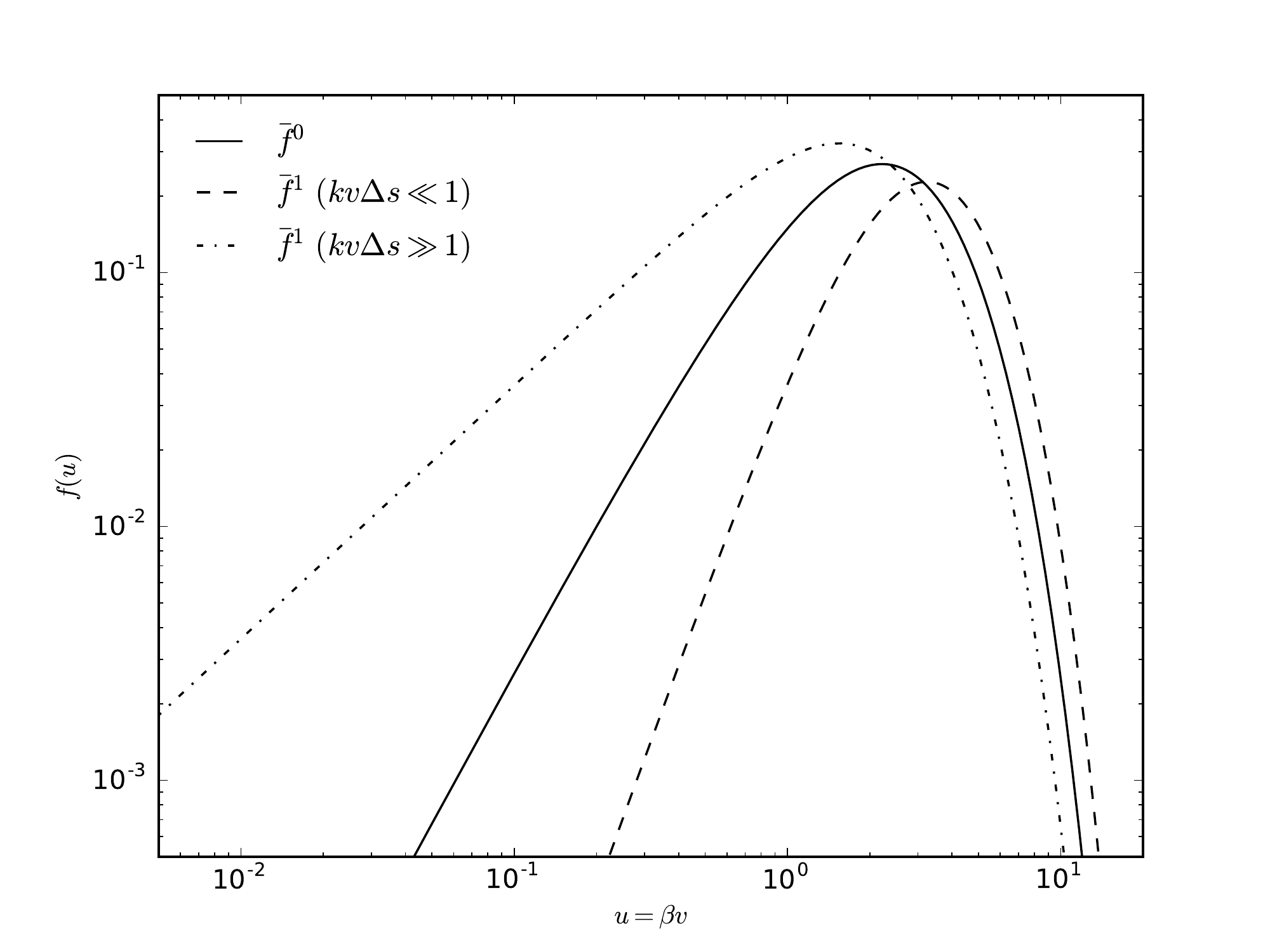}
        \caption{The unperturbed Fermi-Dirac distribution $\bar{f}^0(u)$ as a
          function of $u=\beta v$ is shown as a solid curve.  The
          first order perturbations $\bar{f}^1(u)$ are shown as dashed
          (low-k limit) and dash-dotted (high k limit).  Note that we
          include the $u^2$ part of $d^3u$ in the distributions.
          $\beta=m/(k_BT_\nu c)$}
        \label{fig:distributions}
      \end{center}
    \end{figure}

    We plot $\bar{f}^1(u)$ in Fig. \ref{fig:distributions} and compare
    it to $\bar{f}^0(u)$.  We see that the
    low-k limit tends to shift neutrinos to higher velocities;
    presumably due to the gravitational accelerations.  This is
    qualitatively consistent with the velocity distributions seen in
    simulations, e.g. Fig. 4 and 13 of
    \cite{bib:VillaescusaNavarro2012}.  On the other hand, the high-k
    limit favours low velocity neutrinos and, to our knowledge, has
    not previously been seen.  We have also been unable to find
    particles distributed this way in simple tests of our own
    simulations.

    If neutrinos were distributed according to $\bar{f}^1(u)$ rather
    than $\bar{f}^0(u)$, the asymptotic sound speeds would change.
    For the $kv(\Delta s) \ll 1$ limit, the asymptotic values would be
    $\sqrt{\langle u^2 \rangle_1} = \sqrt{5/3}\sigma$ at low-k and
    $1/\sqrt{\langle u^{-2} \rangle_1}=\sqrt{3}\Sigma$ at high k.  For
    $kv(\Delta s) \gg 1$, the low-k asymptote becomes
    $\sqrt{\langle u^2 \rangle_1} = \sqrt{3}\Sigma$; however, the
    high-k asymptote goes to zero.  This is due to integrating from
    $u=0$, which clearly violates $ku(\Delta s)/\beta \gg 1$ regardless
    of k (the reverse case, for low-k, is less of a problem as
    $\bar{f}^0$ is truncating $u\rightarrow\infty$ and we can also
    simply consider $k=0$).  Hence, the sound speed need not necessarily be
    zero as the approximation technique is somewhat inapplicable.
    Nonetheless, the high-k perturbation is more sensitive
    to low velocity neutrinos and therefore the asymptotic sound
    speed should decrease when including higher perturbations.

  \end{subsection}

  \begin{subsection}{Simulation Sound Speed}
    \label{ssec:simcs}

    Eq. \ref{eq:cs_highk_den} depends only on the total matter density
    field and the neutrino density field.  We can therefore use our
    simulation power spectra to estimate the sound speed with the
    approximation $\delta(k) = \sqrt{\Delta^2(k)}$.  We show the
    results in Fig. \ref{fig:data_cs}.  We find that this estimate of
    the sound speed is significantly lower than the linear theory
    prediction of Eq. \ref{eq:cs_highk} with values $\beta c_s \sim 1$
    rather than $\sim \Sigma$.  There is also now significant mass
    dependence of the asymptotic value, consistent with our
    expectation that heavier neutrinos should behave less linearly,
    and more like CDM (which has no sound speed).

    \begin{figure}[htbp]
      \begin{center}
        \includegraphics[width=0.5\textwidth]{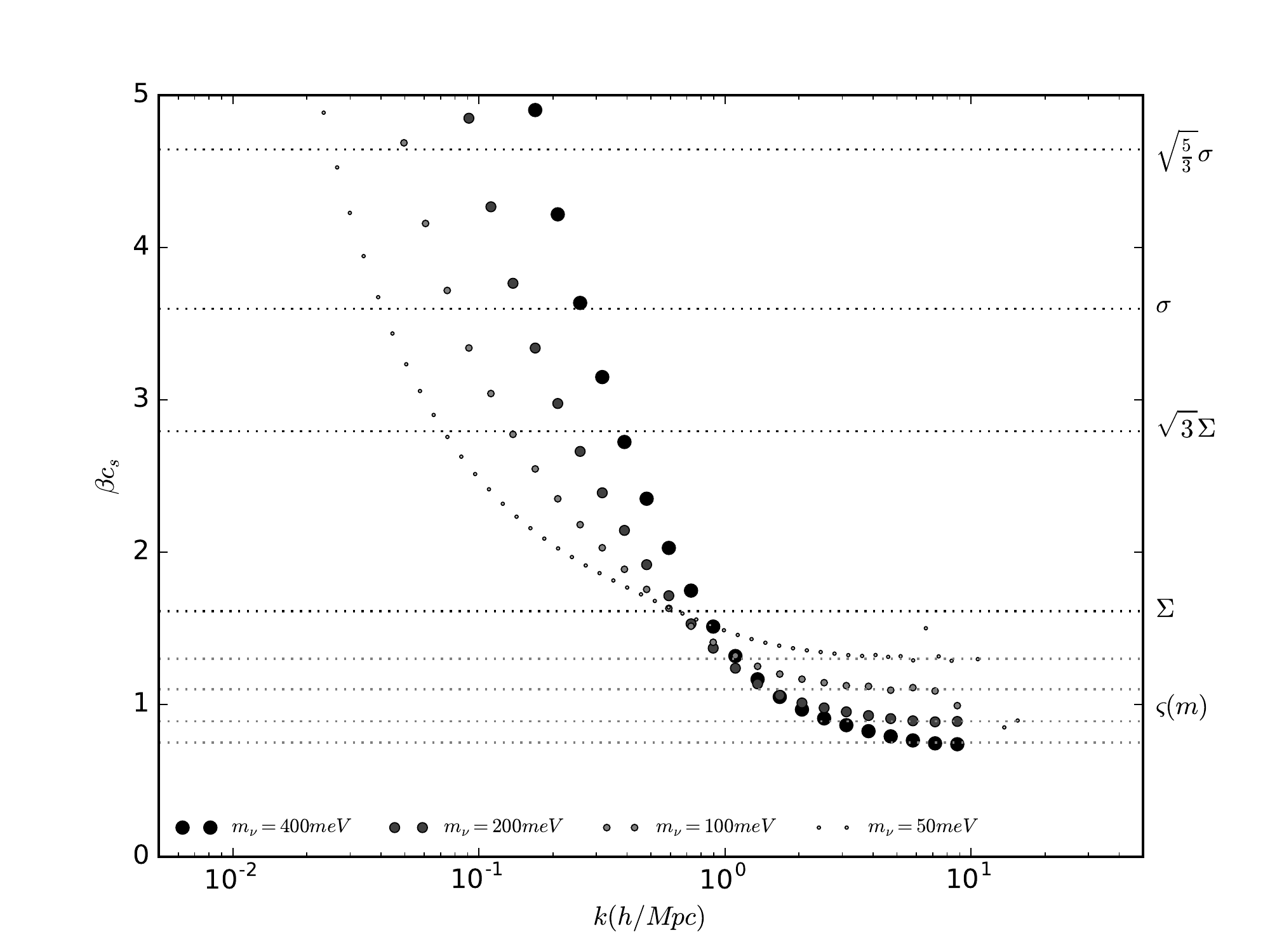}
        \caption{Estimates of the simulation sound speed for different
          neutrino masses.  Dots are estimated using the instantaneous
          approximation (Eq. \ref{eq:cs_highk_den}) and are valid only
          at high k.  Lines show asymptotic values (note that
          $\varsigma$ is calibrated to the instantaneous approximation
          shown).  $\beta=m/(k_BT_\nu c)$}
        \label{fig:data_cs}
      \end{center}
    \end{figure}
    With this behaviour we can now interpret the discrepency between
    linear response and N-body.  The k-dependence of the sound speed
    is proportional to $\delta_m/(k^2\delta)$.  By definition,
    $\delta_m$ is perfectly linear in $\delta_m$.  Therefore, in order
    to drive the sound speed to lower values $\delta$ must be larger.
    This can only occur if, on small scales, non-linearities affect
    $\delta$ much more than they do $\Pi$.  This makes sense as the
    $\Pi$ is weighted by $v^2$ as compared to $v^0$ for $\delta$.
    Hence, we expect Eq. \ref{eq:vlasov_str} to be more accurate than
    Eq. \ref{eq:vlasov_den} as high velocity neutrinos behave more
    linearly.  Furthermore, a decrease in $c_s$ causes the density to
    grow more non-linearly, inducing feedback to continue decreasing
    $c_s$.  Since low mass neutrinos are more linear in $\delta$, they
    are less affected by this instability and so their sound speed is
    closer to the linear theory asymptotic value, $\Sigma$.
    
  \end{subsection}

  \begin{subsection}{Neutrino Power Spectrum}
    \label{ssec:nupower}
    We show the neutrino power spectrum for $m_\nu = 400$ \mev{} in the
    top panel of Fig. \ref{fig:power}.  Black points correspond to the
    N-body results.  Black dashed lines are linear response solutions,
    integrated against linear theory (lower curve) or with a
    non-linear correction:
    $\delta_m \rightarrow \delta_m \sqrt{P_{NL}/P_{L}}$ (upper curve).
    The dashed grey curve is the adiabatic approximation:
    $P_\nu = (T_\nu/T_m)^2P_{NL}$.  We see that it is a reasonably
    good fit to linear response.  However, neither linear response or
    adiabatic solutions reproduce N-body results.  This is despite the
    fact that the usual criterion for non-linearity, $\Delta^2(k) >
    1$, is not met on any scale.  On the other hand,
    the grey curve shows the solution corresponding to
    Eq. \ref{eq:fluid_sln}, with a sound speed measured from
    Fig. \ref{fig:data_cs}, and agrees very well with N-body on small
    scales.  Finally, we show asymptotic behaviour
    (e.g. Eq. \ref{eq:delta_highk}) for linear and non-linear
    potentials as dotted lines.

    \begin{figure}[htbp]
      \begin{center}
        \includegraphics[width=0.5\textwidth]{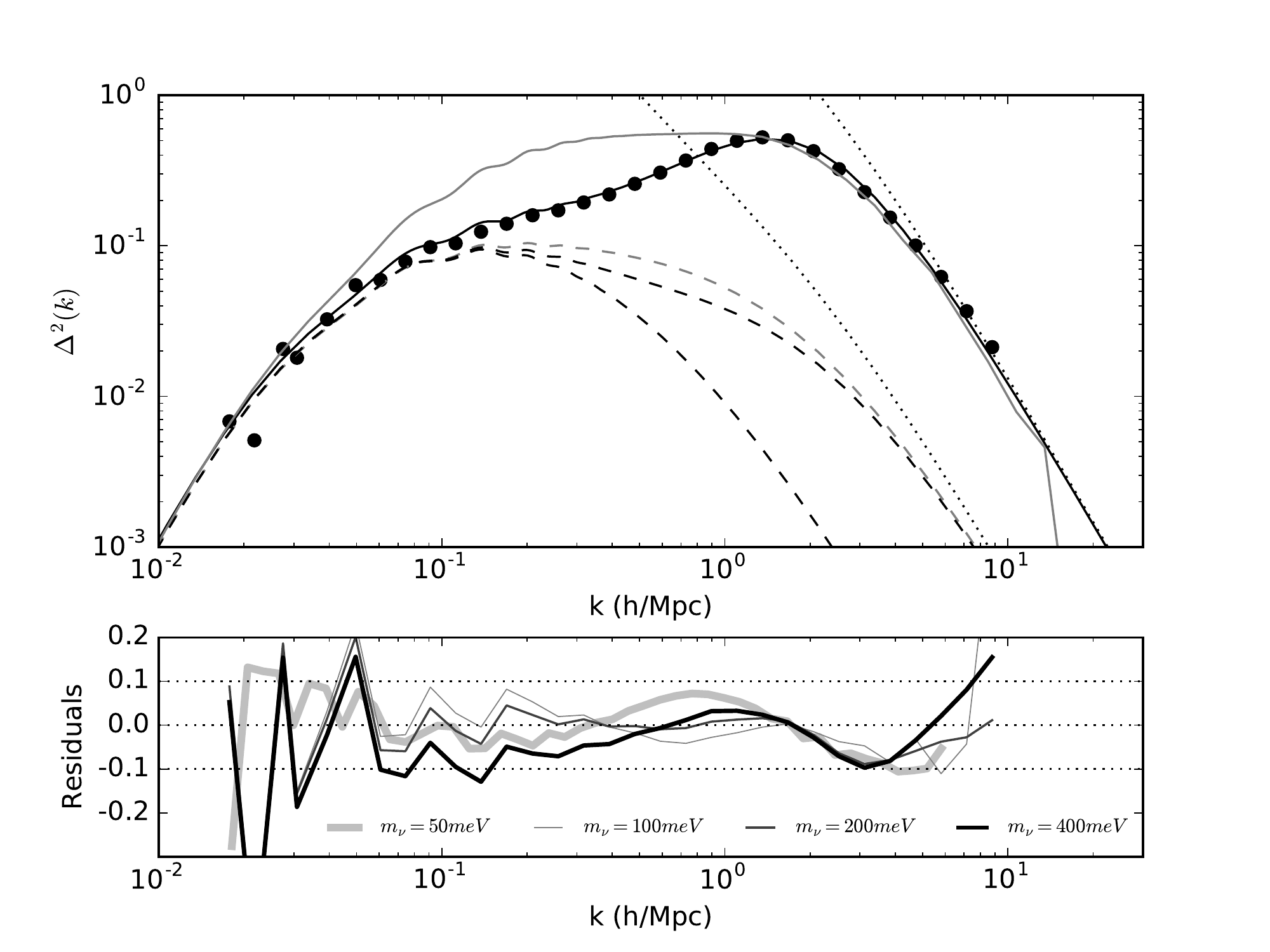}
        \caption{{\it Top Panel} Neutrino power spectrum at $z=0$ for
          $m_\nu=400$ \mev.  Dashed black lines indicate linear response
          to linear (lower curve) and \hfit{} (higher curve)
          potentials.  The dashed grey line is the adiabatic
          approximation.  The dots are from our N-body simulations.
          The solid grey curve corresponds to the sound speed solution
          chosen to match the high k behaviour.  Dotted lines are
          asymptotic behaviours $\propto \delta_m/k^2$.  Finally, the
          solid black line is our model. {\it Bottom Panel} Residuals
          between model and N-body for different neutrino masses.}
        \label{fig:power}
      \end{center}
    \end{figure}

    We now present a simple model relating the neutrino power
    spectrum, $\Delta^2_\nu(k,z,m_\nu)$ to the matter power spectrum,
    $\Delta^2_m(k,z,m_\nu)$:
    \begin{align}
      \label{eq:pmodel}
      \Delta^2_\nu = \Delta^2_m\left[\frac{T_\nu}{T_m} +
      \left( \frac{k_\beta}{k}\right)^2\left( \frac{1}{\varsigma^2}-\frac{1}{\Sigma^2}\right)W(k/k_\varsigma) \right]^2
    \end{align}
    where $T_i$ are linear transfer functions computed via a Boltzmann
    code such as \class{},
    $k_\beta(z) = \Sigma k_{fs} = \sqrt{ \frac{3}{2} \Omega_m a } H_0
    \beta$
    is a typical fluid free-streaming scale neglecting the impact of
    the Fermi-Dirac distribution,
    $\beta c_s = \varsigma=\varsigma(z,m_\nu)$ is the best fitting
    sound speed at high-k, and $W(k/k_\varsigma)$ is a high pass
    filter that truncates the high-k accoustic behaviour.  
    
    We now explain each portion of the model.  The factor of
    $T_\nu/T_m$ corresponds to linear behaviour under adiabatic
    initial conditions (which, as previously noted, is quite close to
    linear response when given the non-linear \hfit{} potential).  The
    second term is the calibrated asymptotic behaviour at high-k, e.g.
    $\delta_\nu \propto \delta_m/k^2$ after subtracting out the linear
    behaviour (with $-1/\Sigma^2$).  We compute
    $\varsigma = \beta c_s$ at redshift $z=0$ by averaging the last
    three points in Fig. \ref{fig:data_cs}\footnote{For $m_\nu=100$
      \mev{} the last point takes a sudden dip so we neglect it and
      average the three points before that.  For $m_\nu=50$ \mev{} we
      average three points in the same k-region as the other masses
      and neglect one point that seems spuriously high.}.  We know it
    must depend on time as it should go to its linear value, $\Sigma$,
    at high redshift.  We compute $\varsigma$ for a few redshifts and
    show the results in the bottom subpanel of
    Fig. \ref{fig:redshift_cs}.
    \begin{figure}[htbp]
      \begin{center}
        \includegraphics[width=0.5\textwidth]{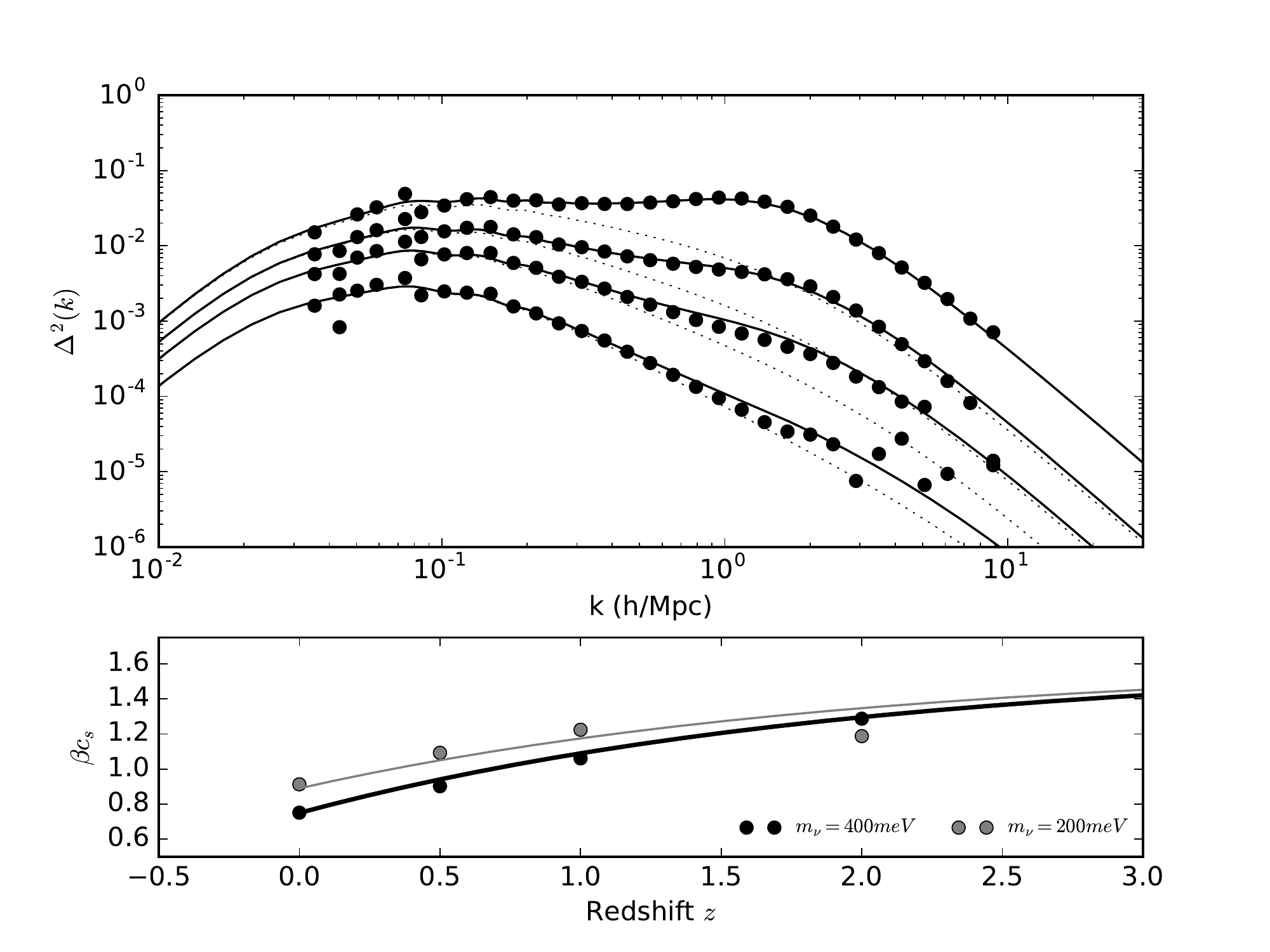}
        \caption{{\it Top Panel} N-body (dots), model (solid) and
          adiabatic (dotted) power spectra at redshifts
          $z=2.0, 1.0, 0.5,$ and $0.0$ for $m_\nu=200$ \mev.  {\it
            Bottom Panel} Redshift dependence of
          $\varsigma = \beta c_s$ evaluated at high k using the
          instantaneous approximation (Eq. \ref{eq:cs_highk_den}) for
          a variety of redshifts and masses.  At redshifts $z>2$ we
          are unable to resolve the value.  Solid lines are for the
          model given in Eq. \ref{eq:redshiftvar}.}
        \label{fig:redshift_cs}
      \end{center}
    \end{figure}
    We model the behaviour as:
    \begin{align}
      \label{eq:redshiftvar}
      \varsigma(z) = \varsigma(0) + (\Sigma - \varsigma(0))Y(z)
    \end{align}
    where $Y(z)$ goes from $0$ at low redshift to $1$ at high
    redshift.  We find $Y(z)=1-e^{-z/2}$ works reasonably well and is
    shown in the bottom panel of Fig. \ref{fig:redshift_cs}.
    Unfortunately we are unable to resolve the power spectra (and
    hence $\varsigma$) at high k for neutrinos below $200$ \mev{} at
    higher redshifts.  Finally, we find the following form for $W$
    provides a good fit:
    \begin{align}
      \label{eq:hpfilter}
      W(x) = \frac{1}{1+x^{-n}}
    \end{align}
    where $n$ must be $> 2$ so that at small scales we recover linear
    behaviour\footnote{We note that filters of the form
      $1/(1+(k/k_{fs})^2)$ have been shown to describe the neutrino
      density contrast quite well in
      \cite{bib:Ringwald2004,bib:AliHaimoud2012} - here we wish to
      model from high-k to low-k so the exponent is negative.}.  We
    find $n=2.25$ allows for good fits to all neutrino masses.  

    Thus, our model has one
    parameter that can be calibrated from simulations,
    $\varsigma(m_\nu, z=0)$, one fitted parameter
    $k_\varsigma(m_\nu)$, and two functions $W(k)$ and $Y(z)$, the
    former depending on $n=2.25$.  We fit between
    $0.2 < k/(\hmpc) < 9 $ (the lower bound is to avoid low-k
    variance) and tabulate these values at $z=0$ in Table
    \ref{tab:varsigma}.
    \begin{table}[h]
      \begin{tabular}{l | c | c | c | c}
        \hline
        Mass (\mev) & 50 & 100 & 200 & 400 \\\hline
        $\varsigma=\beta c_s$ & 1.30 & 1.10 & 0.89 & 0.75 \\ 
        $k_{\varsigma} (\hmpc)$ & 1.11 & 0.92 & 1.20 & 1.65\\\hline
        \end {tabular}
        \caption{Parameters used in modeling the neutrino power
          spectrum.  $\varsigma$ is the dimensionless sound speed calibrated from
          high-k measurements in N-body simulations.  $k_\varsigma$ is a
          best fit parameter.}
      \label{tab:varsigma}
    \end{table}
    This model is shown as a solid black curve in Fig. \ref{fig:power}
    and residuals for all neutrino masses are shown in the bottom
    subpanel.  We see that in regions $0.1 < k/(\hmpc) < 10$ our model
    is accurate to $~10\%$.  Since we do not consider time dependence
    of $k_\varsigma$ (or $n$), at higher redshifts our model does not
    describe the simulated power spectra as well.  For instance, for
    $m_\nu = 400$ \mev{} there is over $ 50\%$ difference at $z=0.5$.
    Nonetheless, this is still much better than linear response, as seen in
    the top panel of Fig. \ref{fig:redshift_cs} where we show power
    spectra at various redshifts along with the adiabatic
    approximation for $m_\nu = 200$ \mev.

  \end{subsection}

\end{section}

%% file: discussion.tex
\begin{section}{Discussion}
  \label{sec:discussion}

  Recently, \citet{bib:Banerjee2016} performed numerical simulations
  treating neutrinos as a fluid.  They evolved the density and
  velocity fields using the continuity and Euler equations and
  estimated the full position-dependent stress tensor from N-body
  neutrino particles.  While this is the most accurate way to close
  the neutrino hierarchy equations, other possibilities exist
  including those discussed here.

  In Fig. \ref{fig:data_cs} we demonstrated that, on small scales, the
  non-linearity in the sound speed is due to the neutrino density, not
  its stress.  In addition, on large scales the behaviour becomes more
  linear and the sound speed is unimportant as
  $j_0(k c_s (s-s')) \simeq 1$.  We speculate that it may be sufficient
  to close the hierarchy equations using an approach analagous to
  \cite{bib:AliHaimoud2012} but using linear response to compute $\Pi$
  instead of $\delta$.

  The benefits of such a scheme are significant compared to N-body.
  For instance, in our particle implementation there are $N^3$
  neutrinos and $(N/2)^3$ CDM particles, each requiring six 4-byte
  floats.  Hence, neutrinos are allocated $8/9$ of the available
  memory.  On the other hand, in a grid based implementation there
  could be $N^3$ CDM particles, and two grids with $(N/2)^3$ cells
  each requiring one 4-byte integer.  In this case neutrinos only
  require $4/100$ of the memory available.  In addition, less
  computational time could be spent on neutrinos (due to the
  simplified hydrodynamic structure) and more time on the CDM.
  Finally, as in \cite{bib:Banerjee2016}, the neutrinos could be
  simulated starting at a high redshift (compared to our N-body
  implementation which starts them at $z\le10$).  If the redshift is
  too high (e.g. above the neutrino relativistic to non-relativistic
  transition), this approach does not accurately describe neutrinos
  (which would be significantly relativistic) but the calculation
  would at least be self-consistent and such high redshift
  discrepencies are unlikely to propagate to late time effects.
  Despite these benefits, a dispersive fluid approach would require
  extensive comparisons to N-body results to calibrate the sound speed
  and validate the results.

\end{section}

%% file: conclusion.tex
\begin{section}{Conclusion}
  \label{sec:conclusion}
  We have considered neutrinos as a dispersive fluid and found that
  this provides additional physical insights into their clustering
  behaviour.  We have computed the sound speed and shown that it
  depends on the intial neutrino velocity distribution and also the
  non-linear cold dark matter.  We find that the excess in power
  observed in the N-body neutrino power spectrum compared to linear
  response can be explained via a higher-order modification to the
  sound speed.  Based on this, we have provided a simple model for the
  neutrino power spectrum that requires no additional integration
  beyond standard Boltzmann code outputs.  Finally, we speculate that
  treating neutrinos as a dispersive fluid could allow for them to be
  simulated efficiently in both memory and processing time.
\end{section}

%% file: acknowledgements.tex
\acknowledgements{
  \label{sec:acknowledgements}
  We acknowledge funding from NSERC.  Computations were performed on
  the General Purpose Cluster supercomputer at the SciNet HPC
  Consortium \cite{bib:Loken2010}.  SciNet is funded by: the Canadian Foundation for
  Innovation under the auspices of Compute Canada; the Government of
  Ontario; Ontario Research Fund - Research Excellence; and the
  University of Toronto.  
}

%% file: supplement.tex
\section*{Supplement}
\label{sec:supplement}
\begin{subsection}{Vlasov Moments}
  \label{ssec:vlasov_moments}
  We describe a simple trick to computing
  $\delta = \frac{ \int d^3v f }{\int d^3v f^0}$,
  $\theta = i k^i \frac{ \int d^3v v^i f } {\int d^3v f^0}$ and
  $\Pi = -k^ik^j \frac{ \int d^3v v^i v^j f } {\int d^3v f^0}$.
  Instead of computing each moment separately, we instead compute a
  ``Moment Generating Function'' - the Fourier transform in velocity
  space of the distribution function.  That is,
  \begin{align*}
    M(s,k,h) = \frac{ \int d^3v e^{-i h^i v^i} f(s,k,v) }{\int d^3v f^0(v)}
  \end{align*}
  where the moments can now be computed by taking derivatives -
  e.g. $\delta = M(s,k,0)$,
  $\theta = k^i \left(\frac{\partial M}{\partial h^i}\right)_{h=0}$,
  $\Pi = k^i k^j \left(\frac{\partial^2 M}{\partial h^i \partial
      h^j}\right)_{h=0}$
  Substituting Eq. \ref{eq:vlasov_sln} in yields:
  \begin{align*}
    M &= \int ds'  a^2 i k^i \phi \frac{ \int d^3v e^{-i v^j (h^j +
        k^j(s-s'))}f^0_{v^j}(v;\beta) }{\int d^3v f^0(v; \beta)} \\
      &= \int ds' a^2 (-k^i g^i \phi) \frac{ \int d^3v f^0(v; \beta)
        e^{-v^jg^j}}{\int d^3v f^0(v; \beta)}
  \end{align*}
  where we integrated by parts and defined $g^i = h^i + k^i(s-s')$.
  The angular part of the velocity integral can be performed
  explicitly by taking the angle between $v^j$ and $g^j$ to be the
  polar angle.  This yields
  $\int_{-1}^1 d\mu_{vg} e^{ivg\mu_{vg}} = 2 j_0(vg)$.  Using this
  result and rearranging yields:
  \begin{align}
    \label{eq:moment_sln}
    M &= \int ds' a^2 (-k\phi) \frac{ \int dv v^2 f^0(v; \beta)
        g\mu_{kg} j_0(vg) }{ \int dv v^2 f^0(v; \beta) } \nonumber \\
      &=\int ds' a^2 (-k\phi) \langle g \mu_{kg} j_0(ug/\beta) \rangle_0
  \end{align}
  where $\mu_{kg}$ is the angle between $k^j$ and $g^j$ and
  $u=\beta v$.  We can immediately find the density in
  Eq. \ref{eq:vlasov_den} by taking $g=k(s-s')$ and $\mu_{kg}=1$.  In
  order to obtain the other moments we need to differentiate.  This
  can be conveniently performed since $g^j \propto h^j$ and so
  $L = k^j \partial/\partial h^j = k^j \partial/\partial g^j$.  In
  spherical coordinates this becomes:
  $L = k \mu_{kg} \frac{\partial}{\partial g} +
  k\frac{1-\mu_{kg}^2}{g} \frac{\partial}{\partial \mu_{kg}}$
  where we use the fact that $M$ only depends on $\mu_{kg}$.  For
  $\theta$ and $\Pi$ this further simplifies since we always take
  $\mu_{kg}=1$ and so we only need: $L=k \frac{\partial}{\partial g}$.
  Applying $L$ and $L^2$ to Eq. \ref{eq:moment_sln} straightforwardly
  gives the velocity divergence, $\theta = i k^i \langle v^i \rangle$,
  and Eq. \ref{eq:vlasov_str}.  In summary:
  \begin{align}
      \delta &= \int ds' a^2(-k^2\phi) (s-s')
               \langle j_0(k u (s-s')/\beta) \rangle_0 \\
      -\theta &= \int ds' a^2(-k^2\phi)
                \langle \cos(ku (s-s')/\beta)
                \rangle_0 \label{eq:vlasov_the} \\
      -\frac{\beta^2}{k^2}\Pi &= \int ds' a^2(-k^2\phi)(s-s')
                                \langle u^2 j_0(k u (s-s')/\beta) \rangle_0.
  \end{align}
  To compute the pressure it is slightly easier to simply start from
  the definition:
  \begin{align}
    3P &= \Pi^{ii} = \frac{ \int d^3v v^iv^i f}{ \int d^3v f^0 } \nonumber \\
       &= \int_{-\infty}^s ds' a^2 ik^i\phi \frac{ \int d^3v v^2 e^{-i
         k^j v^j (s-s')} f^0_{v^i}(v; \beta) }{ 4 \pi \int dv v^2 f^0(v;\beta)} \nonumber \\
       &= \int_{-\infty}^s ds' a^2 ik\phi \frac{ \int d\mu dv v^2
         f^0(v; \beta) (-2v\mu + i k(s-s') v^2) e^{-i k^jv^j(s-s') } }
         {2  \int dv v^2 f^0(v; \beta)} \nonumber \\
       &= \int_{-\infty}^s ds' a^2 ik\phi \frac{ \int dv v^2 f^0(v; \beta) v^2 ( i k (s-s') j_0(kv(s-s')) + 2 j_1(kv(s-s'))/v )}{\int dv v^2 f^0(v;\beta)} \nonumber \\
       &= \int_{-\infty}^s ds' a^2 (-k^2\phi) (s-s') \langle u^2 (j_0(ku(s-s')/\beta) +2 j_1(ku(s-s')/\beta)/(ku(s-s')/\beta) \rangle_0. \nonumber
  \end{align}
  We can now expand $\Pi=(ik^i)(ik^j)\Pi^{ij}$ in terms of the
  pressure and the anisotropic stress: $\Pi = -k^2P + \tau$ and by
  comparison with Eq. \ref{eq:vlasov_str} and the above equation for
  the pressure find
  \begin{align}
          \beta^2 P &= \int_{-\infty}^s ds' a^2 (-k^2\phi) (s-s') \frac{1}{3} \langle u^2
                  (j_0(ku(s-s')/\beta)+\nonumber \\ &\hspace{1cm}+2 j_1(ku(s-s')/\beta)/(ku(s-s')/\beta) \rangle_0 \\
      -\frac{\beta^2}{k^2} \tau &= \int_{-\infty}^s ds' a^2 (-k^2\phi) (s-s') \frac{2}{3} \langle u^2
                                  (j_0(ku(s-s')/\beta) \nonumber \\ &\hspace{1cm}-j_1(ku(s-s')/\beta)/(ku(s-s')/\beta) \rangle_0. 
  \end{align}
\end{subsection}

\begin{subsection}{Fluid Green's Functions}
  \label{ssec:fluid_green_fn}
  Eq. \ref{eq:fluid_eqn} is simply a driven harmonic oscillator:
  \begin{align*}
    \delta_{ss} + \omega^2 \delta = S(s,k)
  \end{align*}
  with $\omega= c_sk$ and $S(s,k) = a^2 (-k^2\phi)$.  We can define a
  Green's function $G(s,s')$ via:
  \begin{align*}
    \delta &= \int ds' G(s,s')S(s')\\
           &= \int ds' G(s,s')(\delta_{s's'}+\omega^2\delta)\\
           &=[G\delta_{s'}-G_{s'}\delta]_{-\infty}^{s} + \int ds'(G_{s's'}+\omega^2G)\delta \\
    \therefore & G_{s's'}+\omega^2G = \delta_{D}(s-s')
  \end{align*}
  and we have the freedom to choose the Green's function boundary
  conditions so as to eliminate the surface term: $G=G_{s'}=0$ for
  $s'>s$.  This has the solution:
  \begin{align*}
    G &= A\sin(\omega s') + B \cos(\omega s') ; s'<s \\
      &= 0; s'>s.
  \end{align*}
  Continuity at $s=s'$ requires $B=-A\tan(s)$ and the jump condition
  $\int_{s-\epsilon}^{s+\epsilon} ds'[G_{s's'}+\omega^2 G =
  \delta_D(s-s')]$
  yields $-G_{s'}(s) = 1$ indicating $A=\cos(ws)/w$ and $B=-\sin(ws)/w$.
  Combining these two yields the Green's function:
  \begin{align*}
    G(s-s') = (s-s')j_0(\omega (s-s'))
  \end{align*}
  and so, for the gravitational source, we find
  \begin{align}
    \delta &= \int ds' a^2(-k^2\phi) (s-s') j_0(k c_s (s-s')).
                   \nonumber
  \end{align}
  $\theta$ is easily computable through $\theta=-\delta_s$:
  \begin{align}
      \theta = \int ds' a^2(-k^2\phi) \cos (k c_s(s-s')). \nonumber
  \end{align}
\end{subsection}